\documentclass[runningheads]{llncs}
\usepackage[T1]{fontenc}
\usepackage{graphicx}
\usepackage{booktabs}
\usepackage[misc]{ifsym}

\usepackage{mwe}
\begin{document}

\title{Music Playlist Captioning at Scale with~Large~Language~Models}

% If the full title of your paper is short enough to also fit in the running head, you can omit the abbreviated paper title here. You can check as follows: if you comment out the \titlerunning line, something will appear in the header of all odd-numbered pages of your PDF from page 3 onward. This something is either the full title (in which case all is well), or the error message "Title Suppressed Due to Excessive Length". If this error message appears, you're going to want to provide an abbreviated title within the \titlerunning command, because if you won't do it, Springer will do it for you.

%N.B.: Author information (both in the \author{} and \authorrunning{} command) should only be present in the Camera-Ready Version of your paper. The version that you initially submit for review, ought to be double-blind. So, when initially submitting your paper, use:
%\author{Author information scrubbed for double-blind reviewing}
\author{Mathieu Delcluze\inst{1} \and 
Léa Briand\inst{1} \and
Benjamin Chapus\inst{1} \and \\ Deniz Mekik\inst{1} \and Guillaume Salha-Galvan\inst{2}}
% You may leave out the orcidID information, if you want to.
% Use \corr to indicate the corresponding author. Note the spacing around the \corr command. Only one author can be the corresponding author.

%N.B.: comment out the \authorrunning{} command for the double-blind version of your paper submitted for review. Later, if your paper is accepted, use the command for the Camera-Ready Version.
\authorrunning{Mathieu Delcluze, et al.}
% If there are two authors, write 'A.L. Benjamin and C.C. Broadus Jr.'
% If there are more than two authors, '[...] et al.' is used.

\institute{Deezer Research 
\and
SJTU Paris Elite Institute of Technology \\ \email{research@deezer.com}}

\maketitle              % typeset the header of the contribution

\begin{abstract}
Music streaming services such as Deezer often recommend personalized playlists to users. Playlist captioning, which involves describing these playlists in natural language, is essential for helping users understand the content behind each recommendation, yet remains challenging at scale. This paper presents the automatic playlist captioning system deployed on Deezer in 2025 to address this challenge. Leveraging recent advances in large language models (LLMs) to generate descriptive captions from diverse data sources in a controlled manner, this system now powers the Daily Mix feature, used by millions of users. This deployment has led to significant improvements in user engagement, highlighting how the semantic framing of an unchanged recommendation shapes user perception in online personalized experiences.
\end{abstract}
\section{Introduction}
\label{sec:intro}

Deezer is a global music streaming service, offering access to over 120 million tracks to 18 million active users across 180 countries~\cite{deezerwebsite}. To help users navigate this vast music catalog, the service relies on large-scale recommender systems, which are essential for driving user engagement and retention~\cite{bendada2023track,bendada2023scalable,bontempelli2022flow,briand2024let,briand2021semi}. In particular, Deezer's homepage features \textit{Daily Mix} playlists, a daily set of personalized music playlists based on each user's listening preferences.

A key challenge with Daily Mix is ensuring that users understand the content behind each recommended playlist. As shown in Figure~\ref{fig:dailymix}, Deezer represents these collections using images combining two album covers along with a short list of selected featured artists. Adding descriptive titles could provide meaningful context. However, summarizing playlists in natural language remains a challenging task, known as playlist captioning~\cite{choi2016towards}. This difficulty is amplified in industry settings, characterized by a large number of playlists.

% and a generic "Daily" title.
%as shown in Figure 1, each playlist was represented by an image combining a few album covers from tracks in the playlist, along with a listing of some featured artists
% and a generic "Daily" title.
%Users would then click on the image to launch the playlist. The use of these generic titles failed to provide meaningful context about the playlist's content.
%Addressing this issue would require describing playlists using natural language to convey their content, a task known as playlist captioning. However, as discussed in Section~\ref{sec:dailymix}, replacing these titles with more informative and representative captions presents significant technical challenges~[CITE]. This is particularly difficult given the large volume of Daily Mix playlists generated on Deezer, as well as the frequent inaccuracies in metadata associated with music tracks~[CITE].

This paper presents the automatic playlist captioning system deployed on Deezer in 2025 to address this challenge. Leveraging recent advances in large language models (LLMs) \cite{epure2025music,wu2024survey}, this production-scale solution generates descriptive playlist captions from diverse data sources in a controlled yet flexible manner. A large-scale online A/B test on millions of Daily Mix users shows significant improvements in user engagement with recommended playlists. Beyond performance gains, our results underscore how semantic framing plays a critical role in shaping user perception of recommendations, even when the underlying content remains the same.

The remainder of this paper is organized as follows. Section~\ref{sec:dailymix} presents the Daily Mix feature in more detail, along with its captioning challenges. Section~\ref{sec:llm} introduces our method, including the motivation for using LLMs, the main components of our pipeline, and the validation and safeguard procedures adopted before deployment. Section~\ref{sec:experiments} reports and discussed our online A/B test results on Deezer, and Section~\ref{sec:conclusion} concludes.

\section{Recommending Daily Mix Playlists on Deezer}
\label{sec:dailymix}

We begin this section by providing a broader overview of the Daily Mix feature, before discussing the challenges of presenting playlist content to users.

\subsection{The Daily Mix Feature}
\label{subsec:dailymixfeature}

The rise of music streaming services has led to the emergence of playlists as a primary way to listen to music \cite{bendada2023scalable,jakobsen2016playlists,schedl2018current}. To help users discover and enjoy content through this format, Deezer offers the Daily Mix feature. This recommender system, prominently displayed on the Deezer homepage, presents five personalized playlists daily to each user, each containing fifty music tracks. These playlists are represented by illustrative covers in a swipeable carousel \cite{bendada2020carousel}, allowing users to easily browse and click to launch them.

Daily Mix playlists are generated based on each user's listening preferences. More precisely, a proprietary algorithm\footnote{Some technical details are omitted in this paper for confidentiality reasons.} combines collaborative filtering and content-based techniques \cite{koren2021advances,schedl2018current} to construct a large graph of similar artists. A Louvain-inspired scalable community detection algorithm \cite{blondel2008fast} extracts hierarchical clusters of similar artists from this graph. Daily Mix then associates users with clusters, considering both (1) relevance, based on user-cluster affinities according to a combination of usage signals, and (2) cluster diversity, using the Maximum Marginal Relevance (MMR) method \cite{carbonell1998use,strasser2025mmr}. 
Finally, Daily Mix generates playlists from each cluster by selecting tracks from artists within the cluster, aiming to balance tracks already familiar to the user with new discoveries.

\subsection{Challenges in Daily Mix Playlist Captioning}
\label{subsec:dailymixlimitation}

\begin{figure}[t]
        \centering
        \includegraphics[width=\linewidth]{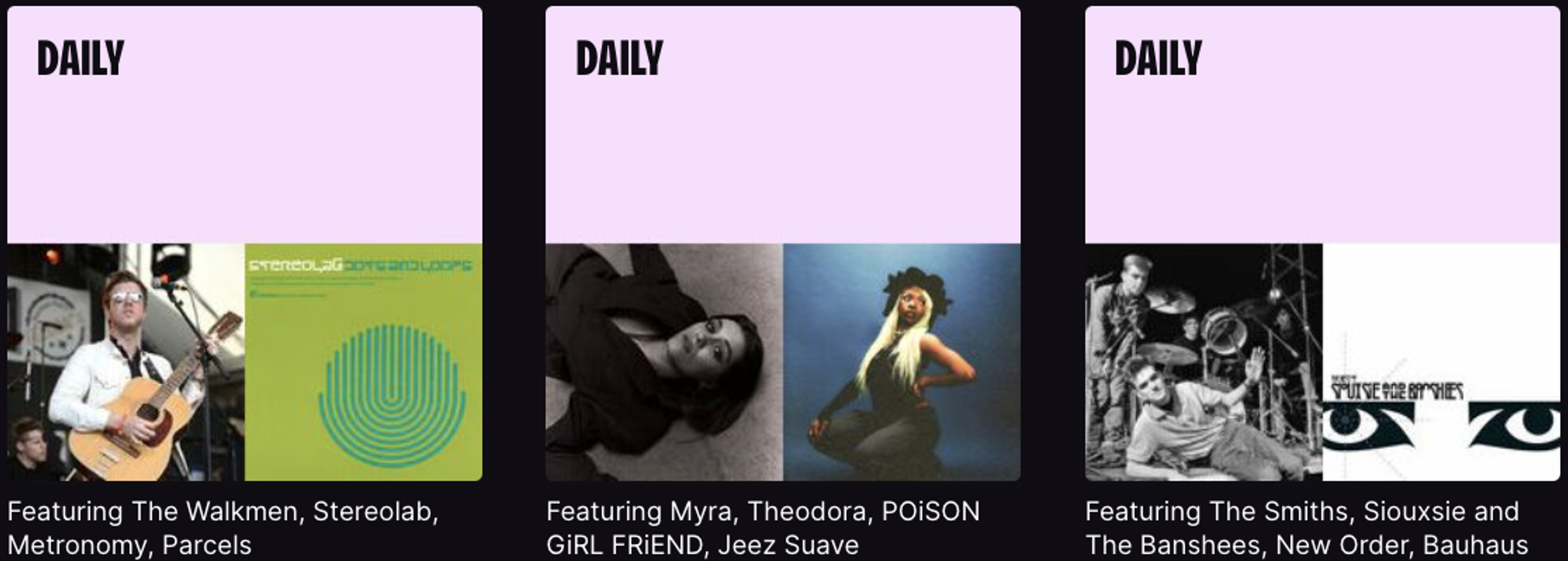}
    \caption{Prior presentation of Daily Mix playlists on Deezer until 2025.}
        \label{fig:dailymix}
\end{figure}

Regardless of recommendation quality, a key challenge with Daily Mix is ensuring that users understand the content of each playlist. As shown in Figure~\ref{fig:dailymix}, playlist illustrations combine two album covers with a short list of featured artists, providing only a rough and often incomplete view of each playlist’s identity. Moreover, the generic "\textit{Daily}" titles are uninformative.

Replacing these titles with descriptive captions, similar to those used for Deezer’s editorial playlists curated by professionals (e.g., "\textit{Modern Rock Essentials}"), could clarify each playlist’s content. However, given the scale and dynamic nature of Daily Mix, providing human-written captions is infeasible. In addition, automatically generating captions remains a technically challenging task, known as playlist captioning. Introduced by Choi et al.~\cite{choi2016towards}, it has since been approached using various deep neural network architectures that aim to model signals such as audio features, linguistic and musicological knowledge, and artist-level side information~\cite{choi2016towards,doh2021music,gabbolini2024surveying,gabbolini2022data,kim23music}.

Nonetheless, prior work still highlights the difficulty of this task. Training data, particularly user-created playlists, is often noisy and lacks consistent templates. A semantic gap also persists when playlist concepts cannot be inferred from track embeddings alone, while linguistic fluency and the inherent subjectivity of captions add further complexity~\cite{gabbolini2022data,kim23music}. Finally, although streaming services often describe music using labels (e.g., genres), which can support captioning, these descriptions remain subjective, incomplete, and sometimes inaccurate~\cite{choi2020prediction,epure2020multilingual,lamere2008social}. Together, these factors make playlist captioning particularly challenging in large-scale industrial settings.

\section{LLM-Based Daily Mix Playlist Captioning}
\label{sec:llm}

In this section, we present our LLM-based method for large-scale automatic playlist captioning in Daily Mix. We first motivate the use of LLMs for this task, to our knowledge for the first time, before presenting our technical pipeline. We then describe a set of caption validation and safety checks adopted internally prior to deployment.

\subsection{Motivations for LLM-Based Captioning}

Unlike previous work, we investigate playlist captioning using LLMs, which are powerful neural models trained on large corpora and capable of generating coherent text from diverse inputs \cite{achiam2023gpt,comanici2025gemini}. This decision is driven by growing evidence of their potential to enhance recommender systems, including by serving as content annotators and summarizers \cite{epure2025music,lin2025can,wu2024survey,zhang2026idproxy,Fan2024RecommenderSI}. Recent work, in particular, shows that LLMs outperform baselines in the related task of music track captioning \cite{bukey2026rethinking,deng2024musilingo,doh2023lpmusiccaps,liu2024music}. We refer to Epure et al.~\cite{epure2025music} for a review of LLM applications and challenges in music~recommendation~specifically.

Our motivation is reinforced by an internal analytics study of user-created Deezer playlists, which revealed significant variability in titles. The tens of thousands of playlists created daily cover a wide range of themes, from genres and artists to periods, occasions, moods, activities, personal events, and ambiguous concepts, with titles varying in length and structure. This diversity underscores the need for a flexible captioning solution, making LLMs~ideal~candidates.

\subsection{Pipeline for LLM-Based Captioning}

\subsubsection{Artist Cluster Tagging for LLM Input}
\label{subsec:tagging} 
\begin{figure}[t]
        \centering
        \includegraphics[width=\linewidth]{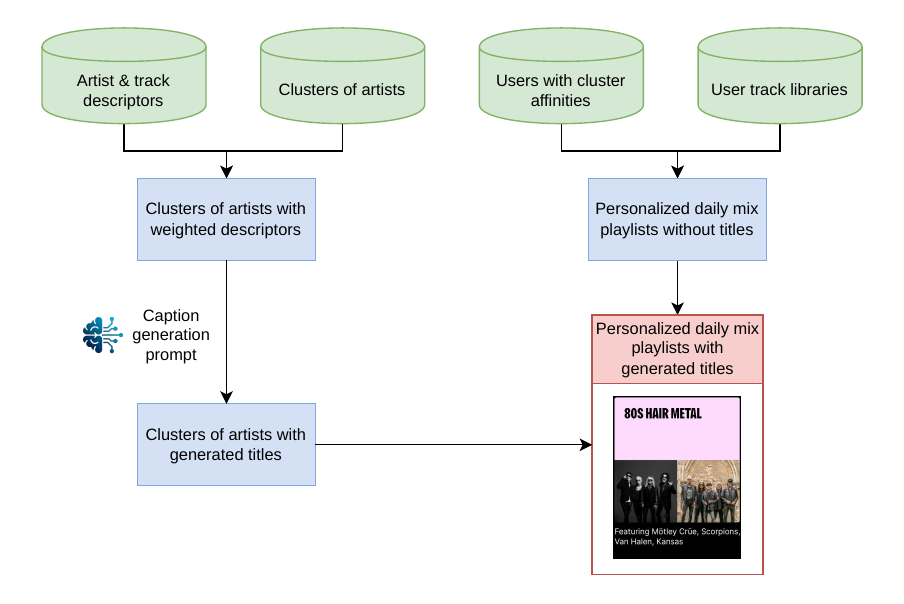}
    \caption{Pipeline overview of music playlist captioning for personalized Daily Mix recommendations in our production environment at Deezer.}
        \label{fig:dailymix-pipeline}
\end{figure}

We now present our technical pipeline, illustrated in Figure~\ref{fig:dailymix-pipeline}. As explained in Section~\ref{subsec:dailymixfeature}, Daily Mix playlists are built from clusters of similar artists. To maintain scalability in playlist captioning, we operate one step upstream of playlist generation and choose to generate captions for the approximately 5,000 existing artist clusters on Deezer, rather than for the millions of personalized playlists produced daily. This approach significantly reduces the computational load in our production environment.

For each cluster, we first associate a range of labels, or \textit{tags}, from Deezer's proprietary catalog, linking these tags to tracks and artists. These tags cover various descriptive aspects of music, including genres \cite{epure2020multilingual,Hennequin2018}, moods~\cite{bontempelli2022flow}, countries \cite{salha2021cold}, and decades. They are obtained through a combination of complementary methods, including audio signal analysis, textual analysis of artist biographies and other music descriptions, and manual annotations by professional editors. Each tag-track pair is assigned a confidence weight.

In addition, artists from each cluster appear in user-created playlists on Deezer. We associate information from titles of these playlists with each cluster, as user-written titles tend to be more creative and original than standard tags, providing a complementary perspective. To this end, we clean the titles by normalizing the text, removing stop-words, and filtering out short or non-informative titles, e.g., those containing only artist names. After cleaning, we compute the frequency and weight of each descriptor associated with tracks and aggregate them at the artist cluster level.

%This produces, for each cluster, a ranked list of weighted descriptors, which is combined with the most representative artist names.

%As explained in Section 2, Deezer daily mixes are directly built from artist clusters mixing representative tracks of the cluster and familiar tracks for the users. To keep daily mix captioning scalable, we operate one step upstream of the tracklist generation itself, at the artist cluster level. This enables us to create captions for only the ~5,000 existing artist clusters, instead of the millions of personalized mixes generated every day.

%Each hierarchical cluster is represented by a group of artists. These artists appear in many user-made playlists, which are typically described by meaningful user-written titles. We clean those titles through normalization, stop-word removal, and for example filtering out short or non-informative titles, excluding titles containing only artist names etc. After cleaning, we compute the frequency and weight of each descriptor associated with an artist, then aggregate them at the artist-cluster level. This produces, for each cluster, a ranked list of weighted descriptors, which is combined with the most representative artist names of the cluster.

\subsubsection{Prompt Design and LLM Captioning}

The steps outlined above associate, for each cluster, a ranked list of weighted descriptors, which we combine with a list of representative artists. This data serves as input for a carefully designed prompt, which instructs an LLM to act as a creative playlist curator.
This LLM initiates a multi-step process, where it first analyzes the input lists, identifies descriptive elements, and cross-checks them with artist information. 

The LLM then generates short and evocative playlist titles in five key languages for Deezer (English, French, German, Portuguese, and Spanish), adhering to formatting constraints specified by product design. 
The prompt ensures that titles focus on specific genres, regions, and thematic elements while filtering out irrelevant or conflicting signals. It also encourages the detection of unifying themes, such as festivals or events, that cohesively link the tracks. Strict rules ensure consistency across languages and maintain a concise format, with titles limited to a maximum of 16 characters. 

%The prompt (given in Appendice A\footnote{Add a link to supplementary material?}) instructs a language model to act as a creative playlist curator, generating short, natural, and precise playlist titles in multiple languages under strict formatting and length constraints imposed by the product design. It relies exclusively on the ranked descriptors (as genre and mood signals) and the representative artists (as semantic validation). The model identifies the dominant sub-genre and key contextual elements, while filtering out noisy or conflicting signals. Beyond genre and mood, it is encouraged to detect additional unifying semantic themes - such as a festival, cultural scene, decade, location, activity, or event - that coherently connect the tracks. Strict formatting rules ensure concise and consistent multilingual titles, while explicit content warnings and safeguards prevent the use of NSFW, offensive, culturally sensitive, or ambiguous terms.

\subsubsection{Deployment} 
The pipeline is designed for seamless integration into large-scale production environments. In practice, we use Spark/Scala for data preparation and Python for artist clustering and playlist captioning. The LLM we employ is Gemini 2.0 Flash~\cite{comanici2025gemini}, chosen primarily for its cost-effectiveness, based on internal benchmarking of various models. Specifically, we generate five candidate captions for each of the 5,000 clusters. To meet production throughput and latency constraints, the LLM API calls are executed in parallel, allowing the entire generation process to complete in under 15 minutes at a nominal cost of approximately \$1~USD. Daily Mix playlists and their captions are generated offline daily and exported to a Cassandra cluster, from which they can be displayed at scale~on~Deezer. 

\subsection{Captioning Validation and Safeguards}
Before being exposed to users, playlist captions had to undergo a series of internal quality and safeguard validation checks. First and foremost, our LLM prompt explicitly prevents the generation of offensive or inappropriate content. An extensive human study involving professional editors from Deezer also evaluated the compliance and faithfulness of captions before exposure to users.

We complemented the human validation with an LLM-as-a-Judge \cite{li2025generation} verification procedure. Another instance of our LLM ran a second prompt, instructing it to act as a music taxonomy and catalog quality expert. Given playlist captions and tags, the LLM assigned quality scores between 0 and 10, reflecting how accurately each caption described the playlist, with an average score of 7.7/10.

A consequence of these checks was that we deployed the model for only three of the five languages (English, French, Portuguese), as the other two did not meet strict quality standards yet. This quality difference may be explained by the overrepresentation of English, French, and Portuguese speakers on Deezer, leading to more user playlists in these languages and thus more  annotations.

\section{Experimental Evaluation} 
\label{sec:experiments}

We now report and discuss our experimental evaluation of the method~on~Deezer.

\subsection{Experimental Setting}
In September and October 2025, we conducted an online A/B test on millions of Deezer users. During this test, a randomly selected cohort received LLM-generated Daily Mix captions, while a control group experienced the original system. 
We compared both approaches using three performance indicators for user engagement: \textit{adoption}, defined as the proportion of users who streamed a recommended playlist at least once during the test; \textit{reconnection}, the percentage of users who streamed in at least three different weeks; and \textit{satisfaction}, the percentage of users who added  tracks to their favorites or a playlist, or exhibited long listening times with low skip rates, according to internal thresholds.

\subsection{Results and Discussion}

Table~\ref{tab:onlineabtest} shows the relative performance of LLM-based captioning compared to the baseline. Adding descriptive captions led to significant improvements in all metrics: a 24.9\% increase in adoption, a 16.9\% increase in reconnection, and an 11.5\% increase in satisfaction (p-value < 0.01 for all metrics). Importantly, we recall that the Daily Mix algorithm was the same for both cohorts; only the captions differed. This highlights how semantic framing can strongly influence user behavior and perception of the same recommendations. In this case, adding descriptions helped users better understand each playlist and drove double-digit engagement gains without modifying the recommender system.

\begin{table}[t]
\centering
% \tiny
\caption{Online A/B test on the Daily Mix feature: Relative improvements of LLM-generated playlist captions vs. baseline (p-value < 0.01 for all metrics).}
\resizebox{\linewidth}{!}{
\setlength{\tabcolsep}{10pt}
\begin{tabular}{c|ccc}
\hline
 & \multicolumn{3}{c}{\textbf{Online A/B Test Metric}} \\ \hline
\textbf{Daily Mix with} & \textbf{Adoption} & \textbf{Reconnection} & \textbf{Satisfaction} \\ 
 \textbf{LLM captioning}   & $+24.9\%$ & $+16.9\%$ & $+11.5\%$ \\  \hline
\end{tabular}}
\label{tab:onlineabtest}
\end{table}

\begin{figure}[t]
    \centering
    \begin{minipage}{0.3\columnwidth}
        \centering
        \includegraphics[width=\linewidth]{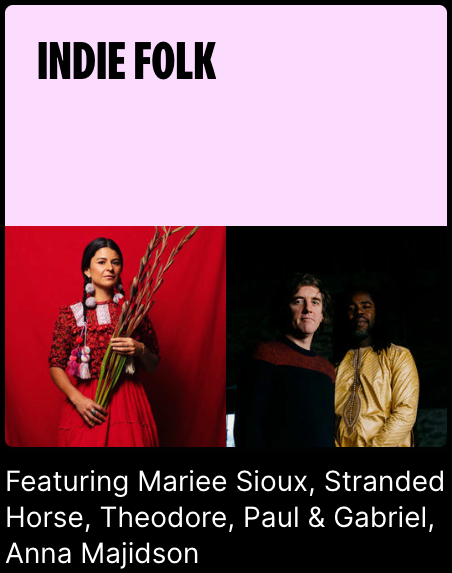}
        %\subcaption{}
    \end{minipage} \hfill
    \begin{minipage}{0.3\columnwidth}
        \centering
        \includegraphics[width=\linewidth]{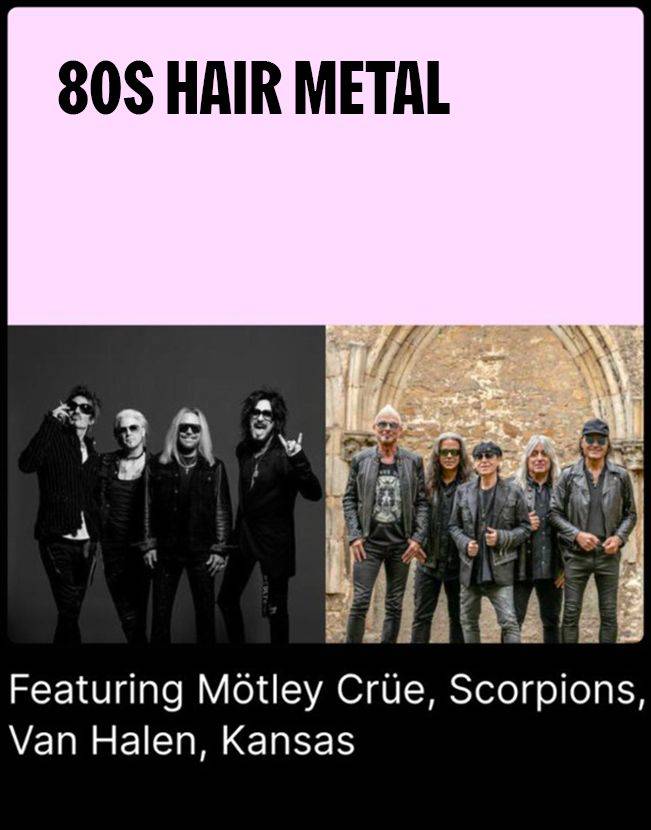}
        %\subcaption{}
    \end{minipage} \hfill
    \begin{minipage}{0.3\columnwidth}
        \centering
        \includegraphics[width=\linewidth]{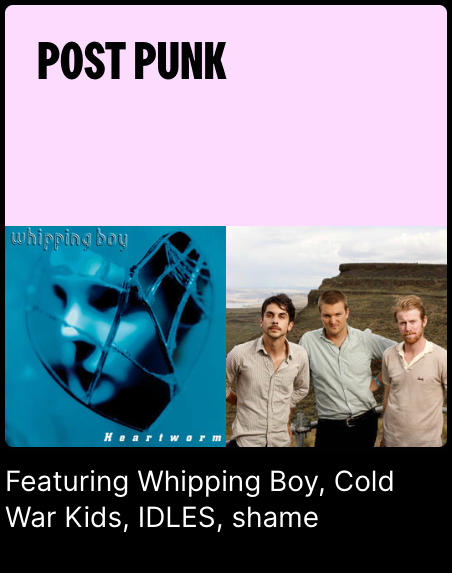}
        %\subcaption{}
    \end{minipage} \\ \vspace{0.5mm} 
    \begin{minipage}{0.3\columnwidth}
        \centering
        \includegraphics[width=\linewidth]{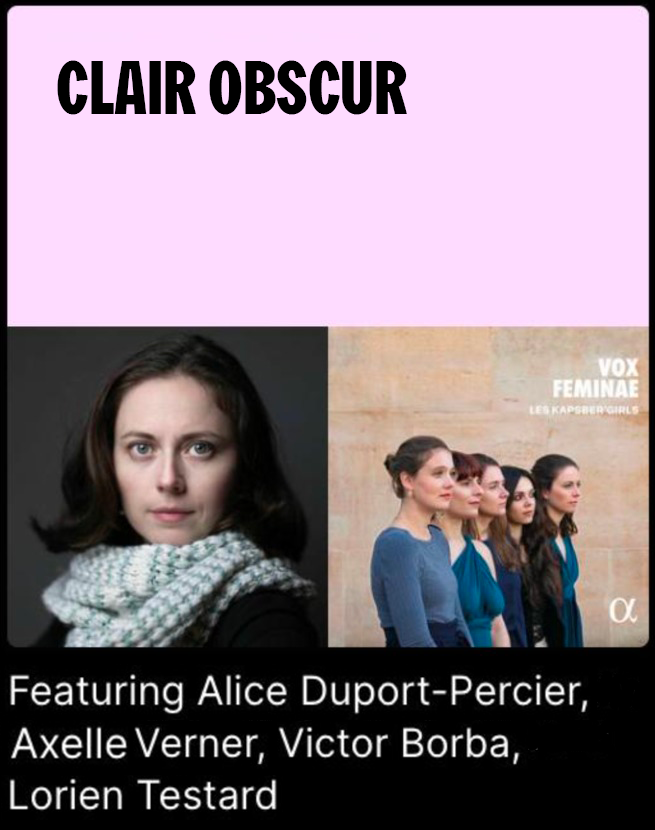}
        %\subcaption{}
    \end{minipage} \hfill
    \begin{minipage}{0.3\columnwidth}
        \centering
        \includegraphics[width=\linewidth]{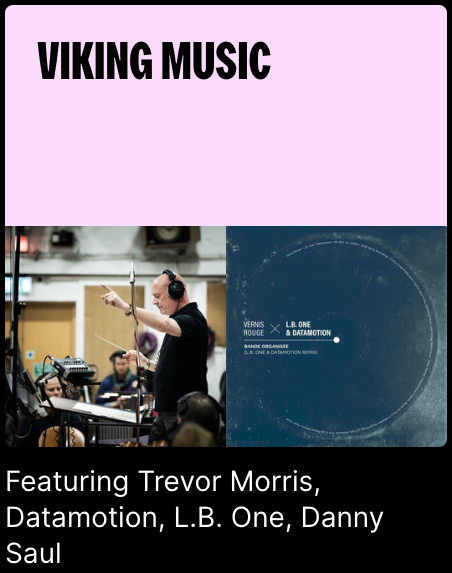}
        %\subcaption{}
    \end{minipage} \hfill
    \begin{minipage}{0.3\columnwidth}
        \centering
        \includegraphics[width=\linewidth]{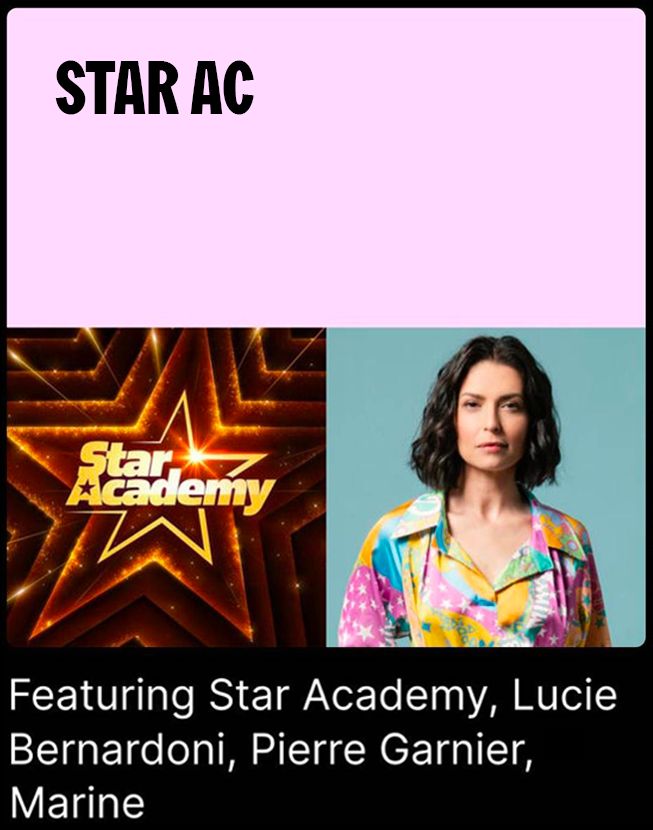}
        %\subcaption{}
    \end{minipage}
    \caption{Examples of LLM-generated Daily Mix captions on Deezer.}
    \label{fig:captionexamples}
\end{figure}

Our LLM-as-a-Judge reveals that, in practice, users are primarily exposed to playlists rated over 8/10, with most scoring 9/10~or~10/10. Popular content on Deezer tends to be associated with richer descriptions during the annotation phase of Section~\ref{subsec:tagging}, resulting in well-graded captions for playlists including this content.
Additionally, Figure~\ref{fig:captionexamples} illustrates that, while the LLM produces genre-based captions (e.g., \textit{Indie Folk}), it also demonstrates flexibility and contextual relevance that goes beyond simple genre-based tagging. For example, \textit{Clair Obscur} accurately describes a playlist of tracks from the video game "Clair Obscur: Expedition 33", while \textit{Star Ac} captions a playlist of artists from a French TV show of the same~name.

In summary, our test highlights the impact of playlist captioning on user perception and engagement, while confirming the utility of our LLM. Its flexibility is a clear strength. However, although no hallucinations were observed, we emphasize that this risk requires careful monitoring. In future work, we will aim to further assess online performance against non-LLM-based methods, benchmark additional LLMs beyond Gemini 2.0 Flash, and improve artist cluster tagging, a key factor for high-quality captions, particularly for less popular content and underrepresented languages.

\section{Conclusion}
\label{sec:conclusion}

In this work, we presented a production-scale automatic playlist captioning system leveraging recent advances in LLMs to generate descriptive captions from diverse data sources in a controlled manner. Through large-scale online A/B testing, we demonstrated significant improvements in user engagement with Daily Mix playlists. These results highlighted the critical role of semantic framing in shaping user perception of recommendations, even when the underlying content remains the same. Our findings also emphasized the flexibility of LLM-based captioning in capturing a wide range of musical themes and contexts. At the same time, our work emphasized the importance of robust validation procedures in real-world deployments, as well as the inherent risk of hallucinations. Based on these positive results, this system has been fully deployed on Deezer since late 2025. It now serves millions of users daily. Looking forward, we outlined several directions to further strengthen our approach. These directions open avenues for future research to further assess the potential of LLM-based captioning.
%% Note that this preceding line implies that you store your BibTeX references in a file called 'mybibliography.bib'. If you instead store your references in a file with a different name, for instance 'references.bib', the preceding line should read '\bibliography{references}'. Whatever you do, DO NOT put the file name extension .bib inside the \bibliography command; this will trip up LaTeX compilers. 
%
% If you do not want to use BibTeX, you can also type up the bibliography exactly as you see fit, using the following structure:

\bibliographystyle{splncs04}
\bibliography{references}

\end{document}